\documentclass[article]{aastex6}
\usepackage{amsmath}
\usepackage{amssymb}
\usepackage{multirow}

\shorttitle{Terrestrial volcanic eruptions}
\shortauthors{Vasilieva $\&$ Zharkova}

\begin{document}
\title{Terrestrial volcanic eruptions and their possible links with solar activity }

\author{Irina Vasilieva}
\affil{Main Astronomical Observatory,  Golosiivo,  Kyiv, Ukraine}
\affil{ZVS Research Enterprise Ltd., London, UK}
\email{vasil@mao.kiev.ua}
\author{Valentina V. Zharkova}
\affil{University of Northumbria, Newcastle upon Tyne, UK}
\affil{ZVS Research Enterprise Ltd., London, UK}
\email{valentina.zharkova@northumbria.ac.uk}

 \begin{abstract}
We compare frequencies of volcanic eruptions in the past 270 years with the variations of solar activity and summary curve of eigen vectors (EVs) of the solar background magnetic field (SBMF) derived from the WSO synoptic magnetic maps. Quartile distributions  of volcanic eruption (VE) frequencies over the four phases of a 11 year cycle (growth, maximum, descent and minimum) reveal higher numbers of eruptions occurring at the maxima of SBMF with southern polarity with some higher levels of eruptions at the minima of sunspot numbers. The frequency analysis of VEs with Morlet wavelet reveals that the period of 22 years to be much stronger pronounced than of 11 years.  Comparison of VE frequencies with the summary curve of EVs of SBMF for 11 cycles after 1868   reveals a strong  positive correlation (coefficient of 0.84) with the southern polarity magnetic fields, while for 8 cycles before 1868 the correlation becomes much lower (coefficient -0.33) and negative. This change of correlation from 1868 can reflect real changes in VE frequencies  caused by migration of the Earth's magnetic pole to lower latitudes, or some differences in proxies of solar cycles. The maxima of VEs  are shown to occur during maxima of solar activity cycles with southern magnetic polarity that can be associated with increased disturbances in the geomagnetic field. The next anticipated maximum of VEs is expected during cycle 26 (2031-2042), when SBMF will have a southern magnetic polarity that can affect solar radiation input to Earth in the current Grand Solar Minimum. 
 \end{abstract}



\section{Introduction} \label{intro}

Volcanos occur when the inner energy of Earth under its surface approaches a certain critical level, so that the hot magma, which is  under a huge inner pressure, can squirt from the Earth's interior through a weak region of the crust producing volcanos. Geomagnetic storms induced by interplanetary coronal mass ejections,  interplanetary magnetic fields and solar wind particles \citep[see, for example][and references therein]{Stamper1999} can cause either sporadic electric currents in the earth locations along the breaks of the surface, which can heat up the surface and reduce their resistivity to the shifts \citep{Han2004}, or induce the currents leading to piezoelectric tension of the breaks  on the surface leading to volcanos \citep{Sobolev1980}.  Magnetic storms occurring during maximum years of solar activity are found to affect the properties of the faults and gestate in some regions with large earthquakes  \citep{Han2004, Love2013}. A clear correlation was detected between the occurrence of large earthquakes at any terrestrial location and high-speed solar wind streams and/or proton densities, which are known increasing during the maxima of solar activity  defined by averaged sunspot numbers \citep{Gonzales1999, Odintsov2006, Marchitelli2020a, Marchitelli2020b}.

The Volcanic Explosivity Index (VEI)  \citep{Newhall_Self1982} was introduced for evaluation of the eruption effects on the terrestrial atmosphere based on the estimation of the volume of volcanic eruption materials (ejected tephra, ashflows, pyroclastic flows etc), the height of the ash column, duration of eruption. The eruptions with the VEI=6 and higher can cause the effect of a volcanic winter - a noticeable cooling of the atmosphere caused by the ash pollution that can, in turn, cause anti-greenhouse effect shielding the solar radiation leading to  global cooling. Therefore,  volcanic activity can be an important component of the solar-terrestrial interaction. 

  The early papers  tried to link the terrestrial volcanic occurrences with solar activity \citep{Kluge1863, DeMarchi1895, Koppen1896, Lyons1899, O'Reilly1898, Jensen1902, Sapper1930}.  Later by applying the wavelet analysis to  the historical records of large volcanic eruptions other authors managed to establish a connection between the global volcanicity and solar activity cycle  of 11 years \citep{Stothers1989, Mazzarella_Palumbo1989}. Other papers  either confirmed \citep{ Casati2014, Ma2018} or denied \citep{Strestik2003} the existence of 11 year cycles in the volcanic eruptions, while   some studies showed close relationships between  earthquakes, solar activity and volcanic eruptions \citep{Schneider1975, Kelly1996}, though questioned by other researchers  \citep{Lockwood2014}. 

This uncertainty exists most likely because there are no  viable mechanisms yet proposed for the explanation of any correlation between volcanic activity on the Earth and solar activity. Most of the authors still support the idea that volcanic activity is increased during minima of solar activity \citep{Stothers1989, Strestik2003, Casati2014} or during a descending phase of solar activity \citep{Herdiwijaya2014}.  The increase of frequency of strong volcanic eruptions during the minima of solar activity was suggested  associated with the variation of circulation of atmospheric masses during a solar cycle \citep{Gray2010}.   Their strong variations are suggested to affect the Earth's revolution about its axis that can cause, in turn, moderate earthquakes and volcanos to relieve the tension of the volcanic magma. This mechanism could reduce a probability of powerful volcanic eruptions while increase the number of moderate eruptions \citep{Stothers1989}. 

\citet{Anderson1974} suggested an alternative model, in which the presence of sufficient quantities of volcanic aerosols can change the circulation of the terrestrial atmosphere  to such the extent that it would change the velocity of the Earth rotation that, in turn, leading to an increase of earthquakes and eruption of powerful volcanos. At the same time,  \citet{Bumba1986} demonstrated possible links of longitudinal distribution of solar magnetic fields with geomagnetic disturbances.

Hence, despite definite inks are not clear yet between volcanic eruptions  and solar activity,  these eruptions can have essential consequences for terrestrial environment by the emergence of volcanic lava and ashes, which  can affect terrestrial atmosphere, its energy exchanges, air quality and living conditions in neighbouring cities. Thus, effects of volcanic eruptions, if their frequencies are noticeable, should be included in one or another ways into any models of the global climate changes.

Although, there was a recent development of finding a new proxy of solar activity, the eigen vectors of the solar background magnetic field  (SBMF) derived with  the principle component analysis (PCA) from the synoptic magnetic maps captured by the full-disk magnetograph of the Wilcox Solar Observatory, US \citep{shepherd14, Zharkova2015}. The modulus of the summary curve of the two principal components  of SBMF fits rather closely the averaged sunspot numbers currently used as the solar activity index \citep{shepherd14, Zharkova2015, Zharkova2022}. The advantage of the new proxy, the summary curve of PCs,  is that it not only provides the amplitudes and shapes of solar activity cycles but also captures  the leading magnetic polarities in these cycles and links not only to the sunspot index but also to various solar flare indices \citep{Zharkova2022}. 

The solar activity was shown to  be defined by the solar dynamo action in the two layers of the solar interior producing two magnetic waves having close but not equal periods of about 11 years. The interference of these two magnetic waves leads to a grand period of about 350-400 years for their amplitude oscillations when the normal magnetic wave (and cycle) amplitudes approach grand solar minima (GSM)  caused by the wave's beating effect \citep{Zharkova2015}.  Such grand periods coincide with well-known GSMs as Maunder minimum (MM), Wolf and Oort and other grand minima \citep{Eddy1976}. In fact, the Sun was shown to enter in 2020 the period of a modern GSM lasting until 2053  \citep{Zharkova2015, Zharkova2020a} similar to the previous GSM. 

Recently, \citet{velasco2021}using the Bayesian algorithm applied to the averaged sunspot numbers obtained the similar results  reporting the modern Grand Solar Minimum to occur in cycles 25-27, similar to that reported by \citet{Zharkova2015}. Furthermore, these prediction results about the modern GSM in cycles 25-27 were confirmed by some other researchers \citep{Kitiashvili2020, Obridko2021}, who used the same WSO synoptic magnetic field data and obtaining the spectra of the zonal harmonics of the SBMF approaching the GSM, which were interpreted with 3D solar dynamo models.  

During the most recent GSM, MM, there was a reduction for solar radiation \citep{Lean1995}  and the terrestrial temperature by about 1C \citep{Easterbrook2016}, which, in turn, was proxied by the absence of sunspots and active regions on the solar surface during the MM \citep{Eddy1976}. Although, the terrestrial temperature  was found increasing since Maunder minimum by 0.5C  per century \citep{Parker1994, Akasofu2010}, which was first assigned to the increase of solar activity  producing a modern warming period \citep{Lockwood1999}.  However, from cycle 21 the solar activity  became systematically decreasing  that coincided with a decrease of the solar background magnetic field in the approach of the GSM \citep{shepherd14, Zharkova2015}. And indeed, from cycle 21 the solar activity  became systematically decreasing  that coincided with a decrease of the solar background magnetic field in the approach of the grand solar minimum (GSM) \citep{Zharkova2015, Zharkova2022}.

On the other hand, in the past few hundred years the Sun was shown to provide some additional radiation to the Earth by moving closer towards the Earth orbit because of the solar inertial motion (SIM) caused by the gravitation of large planets  \citep{Zharkova2019, Zharkova2021}. These periodic variations of the Sun-Earth distance, and the solar irradiance, occur every 2100-2200 years, called  Hallstatt's cycles, which were independently derived from the isotope abundances in the terrestrial biomass  \citep{Steinhilber09, Steinhilber12}. In the current Hallstatt's millennial cycle, the Sun-Earth distances are decreasing from the MM until 2600 that leads to the increase of solar irradiance deposited to the atmosphere of the Earth (and other planets)  \citep{Zharkova2021}. 
This SIM effect is likely to contribute to the terrestrial atmosphere heating, in addition to any heating caused by the greenhouse gasses considered in the terrestrial models that requires further investigation. 
 
 However, the most essential effect of SBMF in the next few decades will come from a reduction of solar activity, or the modern grand solar minimum, which started in 2020 and will last until 2053 \citep{Zharkova2015}. Since geomagnetic storms can be associated with volcanos \citep{Sobolev1980}  either by causing sporadic electric currents \citep{Han2004}, or by inducing the currents leading to piezoelectric tension of the breaks  on the surface, the scope of the current paper is to establish  more definite  links between  the frequency of volcanic activity and the variations of solar activity using both the sunspot numbers \citep{ Wolf1870} and the new proxy of solar activity linked to the solar background magnetic field \citep{Zharkova2015}.

\section{Observations} \label{distil} \label{obser}
\subsection{Frequency of volcanic eruptions} \label{volc_freq}
The information about the volcanic eruptions was derived from the Smithsonian Institution's GVP database including the volcanoes that are known or suspected to have erupted within the Holocene or Pleistocene. The GVP website provides access to the raw data and a history of volcanic eruptions. The main list of Holocene volcanoes contains 1408 volcanoes associated with 9928 eruptions (version 4.10.0 dated May 14, 2021) \citep{Global2013}. The dates of volcanic eruptions are determined by different methods and with varying precision. There are 2 criteria, which a volcanic eruption must satisfy, in order to be included into this research:  it must be accurately dated, and it must be historically confirmed. 

\begin{figure}[h]
\center{\includegraphics[width=40pc]{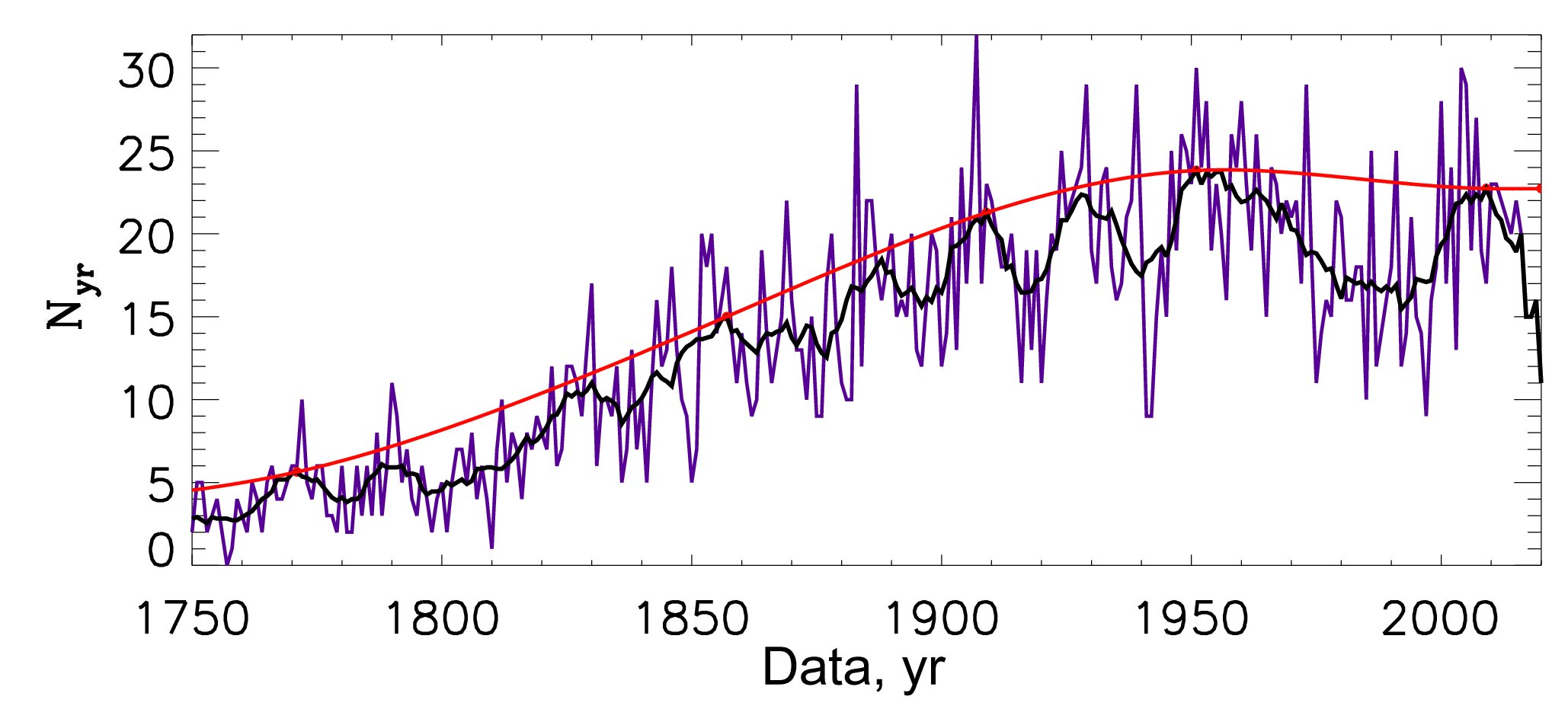}}
\caption{Frequencies of the annual volcanic eruptions, $N_{yr}$,  in the period of 1700-2020 (violet line). The annual numbers of  volcanic eruptions averaged with a running filter of 11 years (black line) over-plotted with the  envelope curve (red line) marking maximal magnitudes  in the black curve.}
\label{volc_erup}
\end{figure}

Fig. \ref{volc_erup} shows the annual frequency of volcanic eruptions, $N_{yr}$, that were recorded from 1700 to 2020 (5639 in total, of which 3943 with VEI $\ge 2$). In the last century, the frequency of observed volcanic eruptions, $N_{yr}$,  is higher, and in the twentieth century (1900-1999), 2944 eruptions were observed (of which 1940 with VEI $\ge 2$). However, the fixation of a larger number of the eruptions in recent years is most likely due to changes in the method registrations of the eruptions. According to \citep{Herdiwijaya2014} completeness of the dates of volcanic eruptions for VEI $\ge 5$ begins since 1800, for VEI $\ge 4$ - since 1900 and for VEI $\ge 3$ - since 1960. In our analysis, we are assumed that the data gaps are a random process, and the trend in the data has been eliminated by the spline. 

In order to eliminate random variations of the volcanic eruption frequencies,  $N_{yr}$, the data was smoothed with the running averaging filter over an 11 year window and plotted as averaged (black) curve in Fig. \ref{volc_erup}. By assuming that in the past three centuries the maximal number of volcanic eruptions was similar to that  in the past few decades, we have built the envelope curve (red curve) along the maximal magnitudes of averaged curve of eruption occurrences (black curve) and carried out the interpolation by spline for the intermediate magnitudes (Fig. \ref{volc_erup}).


\subsection{Eruptions frequencies versus Wolf numbers}
Solar activity is currently defined by the average number of sunspots and groups on a solar disk at a given day or month. \citep{ Wolf1870}. There is a well-known 11-year cycle of solar activity \citep{Wolf1870}, or a 22-year cycle, during which a complete reversal of the magnetic polarity of the sunspots occurs.  The annual numbers of sunspots were taken from the Solar Influences Data Analysis Center (SIDC) at the Royal Observatory of Belgium \citep{SILSO}.  

\begin{figure}
\gridline{\fig{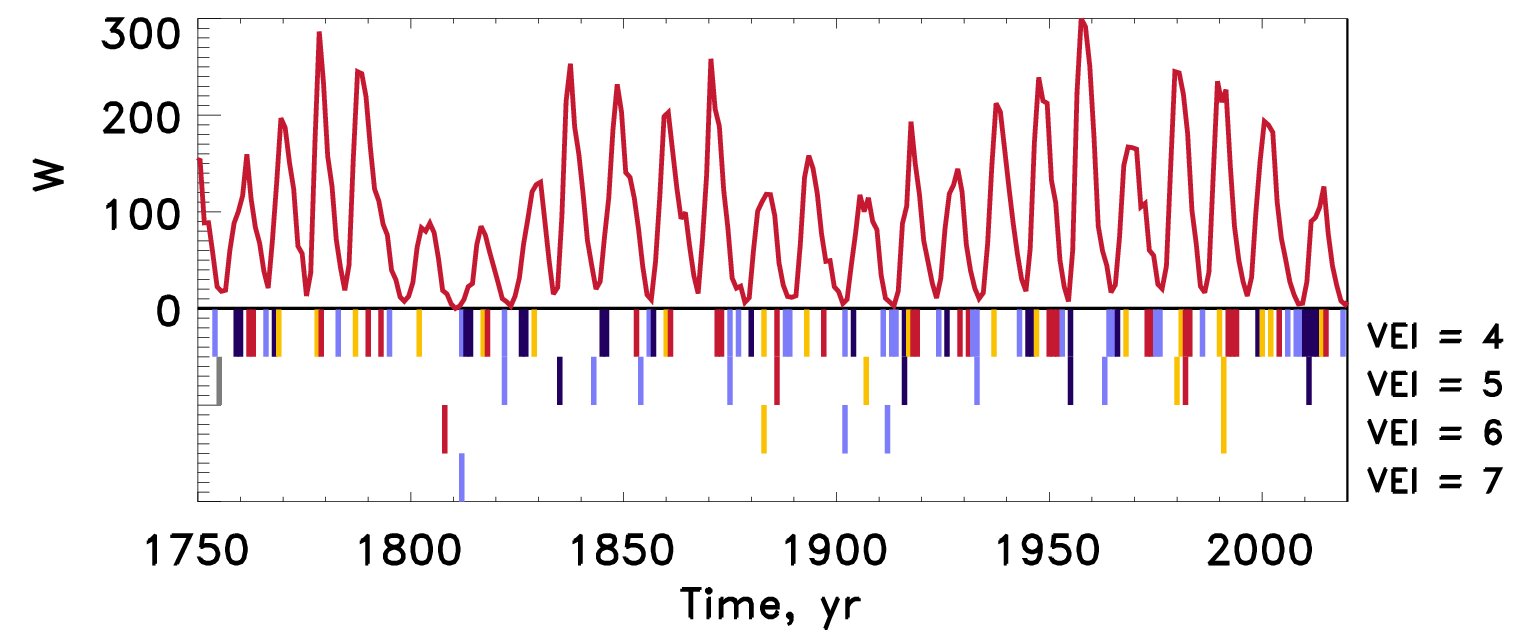}{0.9\textwidth}{a - volcanic eruption (VE) frequencies versus Wolf numbers;}}
\gridline{\fig{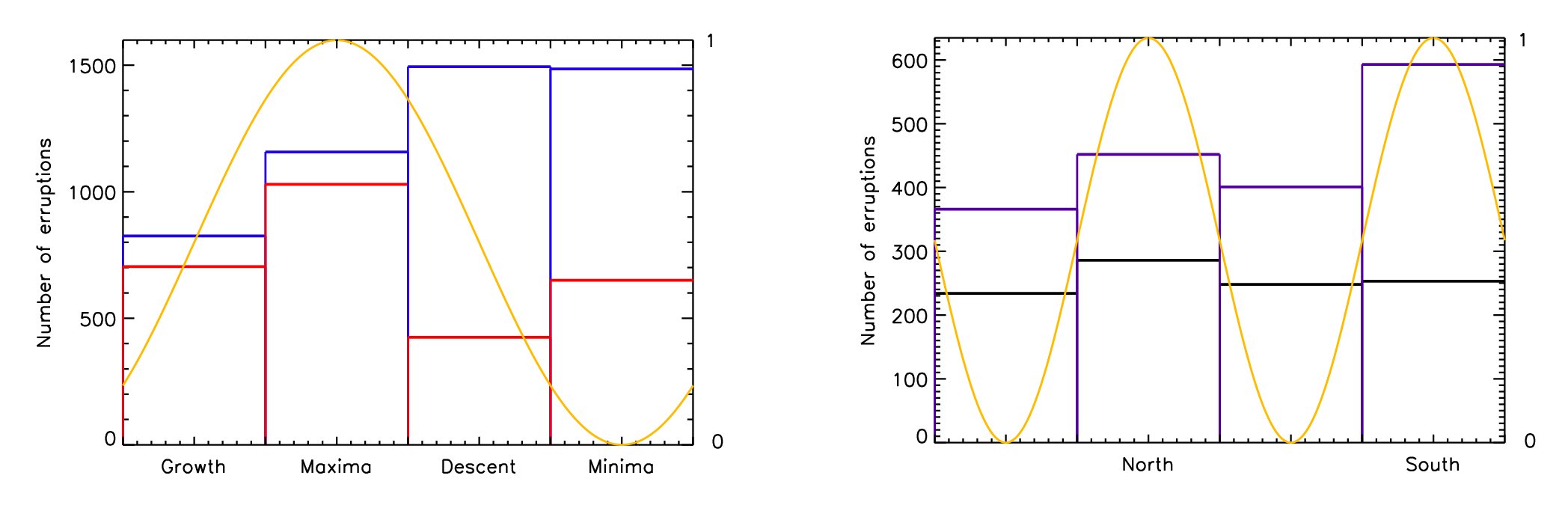}{0.9\textwidth}{b - VE frequencies vs quartiles of 11y-solar cycle (X-axis); \hspace{0.1cm}  c - VE frequencies vs quartiles of 22y-double solar cycle (X-axis). }}
\caption{ a - Wolf numbers (red curve) describing solar activity in 1750-2020 (top) versus the annual frequencies for volcanic eruptions  of different significance  (from lower (top line)  to higher (bottom line) marked by colour bars. Colours  of the bars define the phases of solar activity cycles when eruptions occurred: blue - ascending, yellow - maxima, red - descending and light blue - minima. b - Volcanic eruption frequencies  (blue lines) during the four quartiles of a symbolic 11 year solar cycle (cosine function, yellow line) defined by sunspots (blue line) and by SBMF (red lines). c - Volcanic eruption (VE) frequencies  versus a symbolic 22 year solar cycle (two cosine functions (yellow line) with with the opposite polarities as shown in Fig. \ref{sa_sbmf}, bottom plot), defined by a summary curve of eigen vectors (EVs) of SBMF (Fig. \ref{sa_sbmf}, top plot). The summary VE frequencies were calculated for the intervals marked by rectangular in X-axis showing the four phases of solar cycles. For the EVs  cycles the rectangles in axis X show the intervals of maxima and minima of double cycle  and the leading magnetic polarities in given 11 year cycle (northern  with a positive sign and southern with a negative one) for the two periods: 1750-1868 (black line) and 1868-1950 \&1990-2020 (indigo line).}
\label{wolf_num}
\end{figure}

The  25 cycles of solar activity occurred during the period of 270 years from 1750 until 2020 are shown  in Fig. \ref{wolf_num}, top plot.  The averaged sunspot numbers are plotted versus the available frequencies  of the eruption of volcanos of different significance. It is seems that the very strong volcanic eruptions (VEI $\geq 5$) occur during the minima of solar activity defined by the sunspot index.The further analysis of these links is presented in section \ref{sa_links} below.

\section{Data analysis} \label{disc}
\subsection{Links of volcanic eruptions with the solar activity phases}  \label{sa_links}
 In order to compare volcanic frequencies at different phases of solar cycles, let us us the following phase dates  taken from \citet{Vasilyeva2020}.  For each solar cycle  of 11 years the maximal and minimal numbers of sunspots are calculated, and the temporal interval between them was divided by 8 to obtain smaller intervals, so that each cycle with two minima and one maximum has 16 points. Then for the ascending (growth) and descending  (descent) phases we used 4 intervals, for maximal phase we use also four intervals (two from each side from the maximum) and for minima four intervals (two from the each minima). For the EVs  cycles of 22 years the rectangles in axis X show the intervals of maxima and minima of double cycle (6 small intervals each for 11 year cycle)  and indicate the leading magnetic polarities in a given 11 year cycle (northern  with a positive sign and southern with a negative one).

The distributions of volcanic eruptions in the four quartiles of a solar cycle (yellow line) shown in axis X: growth, maxima, descent and minima using  the rule described in the paragraph above are presented in Table 1 and Fig. \ref{wolf_num} for the whole volcanic dataset (bottom left plot) excluding only 1950-1990 as likely to be affected by the open nuclear weapon testing). The symbolic cycle  of 11 years is represented by yellow cosine function with the Y-axis on the right showing maximal magnitude of 1 (unity)  and minimal magnitude of 0 (zero). The largest volcanic eruptions are found to occur during the descending and minimal phases of the solar activity defined by sunspots that agrees with the previous findings \citep{Stothers1989, Strestik2003, Casati2014, Herdiwijaya2014}. 
 
 However, the frequencies of volcanic eruptions (VEs) can be also compared with the solar activity cycles provided by the summary curve of the eigen vectors (EVs) of SBMF for maximal and minimal quartiles (axis X) for 11 year cycles with the opposite magnetic polarities  similar to those shown in Fig. \ref{sa_sbmf}, top plot,  or their modulus summary curve replicating the sunspot index as shown in Fig. \ref{sa_sbmf}, bottom plot \citep{Zharkova2015}.  These VE frequencies compared for the phases in an 11-year symbolic cycle are shown in Fig. \ref{wolf_num}, bottom left plot (red lines), which demonstrate that the maximal frequencies of VEs occur during the maxima of the solar magnetic cycle defined by the EVs proxy. 

 Furthermore, we  also consider a symbolic double cycle of solar activity of 22 years comprising two 11 cycles e.g. the MSC for cycles 21 and 22 in Fig. \ref{sa_sbmf}, bottom right plot, where cycle 21 has EVs of the northern polarity and cycle 22 has EVs of the southern polarity.  The magnitudes of these symbolic cycles  are shown in the right Y-axis varying from 1 (maximum) to 0 ( minimum) for 11 year cycles.

  In this case  of double cycles there is a clear indication that for the period of 1868-1950 the VE frequencies (indigo lines in Fig.\ref{wolf_num}, bottom right plot) are maximal during the magnetic cycle having the EVs with southern magnetic polarity (the second 11-year cycle in Fig.\ref{wolf_num}) compared to a smaller maximum occurring during the maximum of magnetic cycle with northern polarity (the first 11-year cycle). This finding is closer to the other studies of volcanic and earthquake occurrences in the recent century\citep{Gonzales1999, Odintsov2006, Marchitelli2020a, Marchitelli2020b},  showing the maximal frequencies of volcanic eruptions occurring during the maxima of solar magnetic field activity.   
 
  While for the earlier period of 1750-1868 there is some small increases of VEs  (black line) during the maxima of magnetic cycles with either northern or southern polarity. This finding can be linked either to specific conditions occurred  either at Earth or Sun during this early period of solar activity. Alternatively, it can be caused  by the differences between solar activity curves derived from a small number  of sunspot observations in 18th century that is discussed in the forthcoming paper \citep{Zharkova2022b}.  

However,   in the both definitions  of solar activity (sunspot and magnetic cycles), the number of volcanic eruptions  in the first period 1750-1868 is equally high during the 11 cycle minima as they are during their maxima, e.g. not being affected  by the solar activity indices. This leads us to a conclusion that there could be also some natural terrestrial phenomena (e.g. a north pole migration) causing these steady VE frequencies as discussed in section \ref{n_pole} below.

\begin{table*}
	\centering
	\caption{The total number of volcanic eruptions N  occurred for a given phase of  a solar cycle and $\%$ from the total number of eruptions in the cycle derived  for several VEI in the period of 1750-2020. } 
	\label{tab:solarphases}
	\begin{tabular}{l|c|c|c|c|c|c|c|c|c} 
		\hline
 \multirow {2}{*}{VEI} &  \multicolumn{2}{c|}{Growth} & \multicolumn{2}{c|}	{Max} & \multicolumn{2}{c|}{Descent} & \multicolumn{2}{c|}	{Min} & 
 {All} \\

\cline{2-10}
  &	N &	\% & N & \% & N & \% & N & \% & N \\
	\hline
	1 & 171 & 15.0 & 269 & 23.6 & 324 & 28.4 & 375 & 32.9 & 1139 \\
2 &	523 & 17.0 & 723 & 23.5 & 961 & 31.3 & 864 & 28.1 & 3071 \\
3 & 99 & 16.4 & 140 & 23.3 & 169 & 28.1 & 194 & 32.2 & 602 \\
4 & 27 & 22.4 & 20 & 15.9 & 37 & 29.4 & 42 & 33.3 & 126 \\
5 & 5 & 29.4 & 3 & 17.6 & 2 & 11.8 & 7 & 41.2 & 17 \\
6 & 0 & 0 & 2 & 40.0 & 1 & 20.0 & 2 & 40.0 & 5\\
7 & 0 & 0 & 0 & 0 & 0 & 0 & 1 & 100. & 1 \\
	\hline
All & 825 & 16.7 & 1157 & 23.3 & 1494 & 30.1 & 1485 & 29.9 & 4961 \\
		\hline
	\end{tabular}
\end{table*}

\subsection{Periods of volcanic activity from the wavelet analysis} \label{period}

If the volcanic activity has a periodicity and the dominant one is about 11 years then this would be a good verification of the connection between solar and volcanic activity. This point was investigated using a wavelet transform  \citep{Mazzarella_Palumbo1989}. 

Wavelet transform of signals is the development of spectral analysis methods. Unlike Fourier analysis, the wavelet transform gives a two-dimensional scan of the analysed signal, while the time coordinate and frequency are independent variables. This representation allows you to explore the properties signal simultaneously in time and frequency spaces. Wavelet analysis is excellent a tool for examining signals with time-varying frequency characteristics. 

\begin{figure}[h]
\center{\includegraphics[width=40pc]{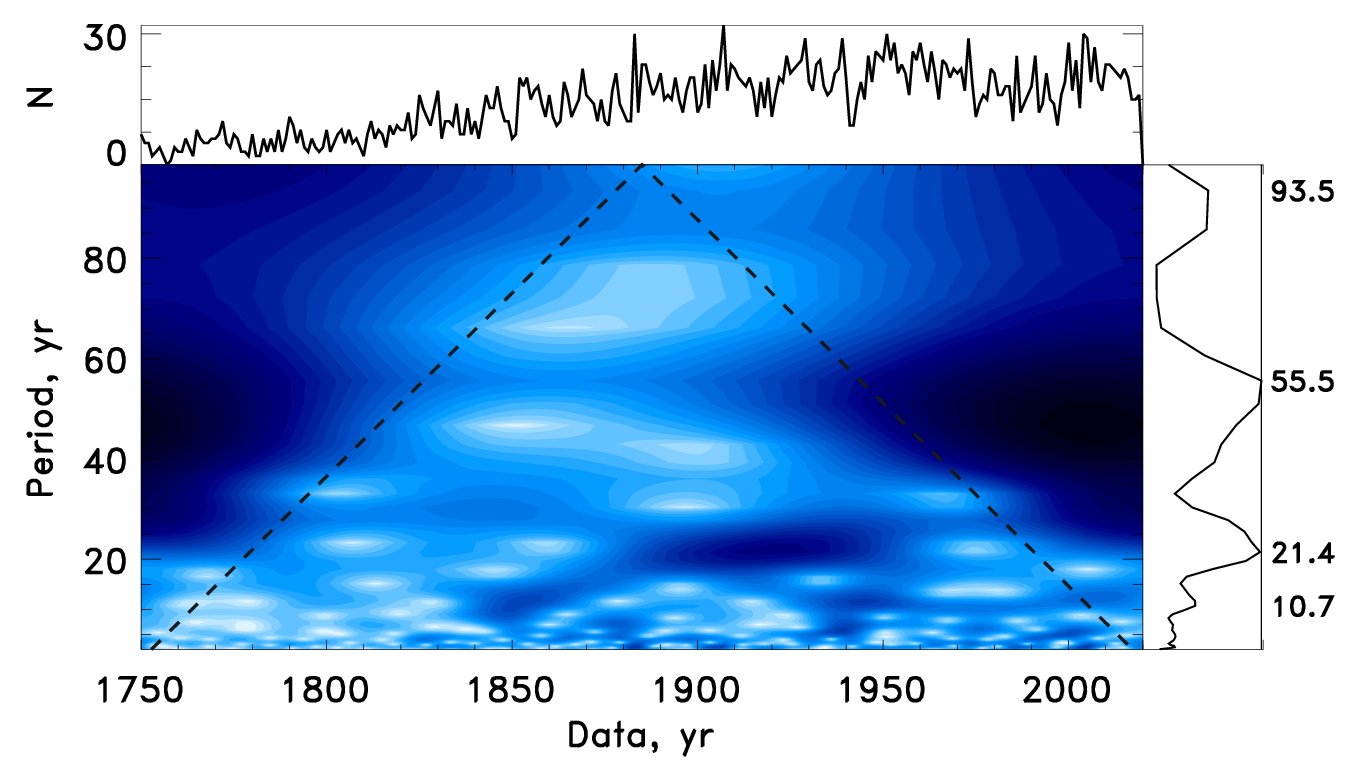}}
\caption{Time series of the annual volcanic eruptions in the period of 1750-2020 (top). The wavelet power spectrum, using the Morlet wavelet. The x-axis is the wavelet location in time. The y-axis is the wavelet period in years. The cone of influence given by the dashed line. The global wavelet spectrum and the main periods of volcanic eruptions (right).}
\label{wavelet}
\end{figure}

The choice of the mother wavelet is dictated by the task and the nature of the signal under study. When we use the Morlet wavelet (the real part of it is damped function of cosine), we obtain a high frequency resolution, which is important for our task.
By considering the time series in  the frequency-time space it is possible to derive dominant periods and their variations in time. 
The Morlet wavelet analysis  was applied to the frequencies of volcanic eruptions reported in section \ref{volc_freq} and the result is plotted  Fig. \ref{wavelet}. The  wavelet spectrum of the temporal series of volcanic eruptions in 1750-2020 years  reveals the two powerful peaks at 21.4 $\pm$ 1.4 years (corresponding to a double 11 year cycle) and at 55.6$\pm$ 10.5 years (its nature is unknown yet). 
 At the same time,  the peak near the period of 10.7$\pm$0.9 (close to the duration of a single solar activity cycle) is found to be much less pronounced. 

This double cycle feature motivates us to investigate the link of volcanic eruptions with the solar activity proxy  expressed in variations of the eigen vectors of SBMF, which covers a 22 year cycle as it contains also a magnetic polarity of SBMF for each cycle.

 \subsection{Volcanic eruption frequency and solar background magnetic field} \label{sbmf}
 \subsubsection{Solar background magnetic field as a new solar activity proxy} \label{sbmf_met}
Recently, the Principle Component Analysis  was applied to the low-resolution full disk solar  background magnetic field  (associated with the poloidal magnetic field) measured by the Wilcox Solar Observatory to derive the dominant eigenvalues covering the maximum variance of the data \citep{Zharkova12, Zharkova2015, Zharkova2022} corresponding to the eigenvectors, EVs, or Principal Components (PCs), which came in pairs. 

These PCs are considered to be a reflection of the main (dipole) dynamo waves in the solar poloidal magnetic field produced by the dynamo mechanism \citep{Zharkova2015}. The PCs were classified by applying  the symbolic regression approach based on the Hamiltonian principle \citep{Schmidt09} and deriving the mathematical formulae describing the amplitude and phase variations \citep{shepherd14, Zharkova2015}. 
 \begin{figure}
\includegraphics[width=35pc]{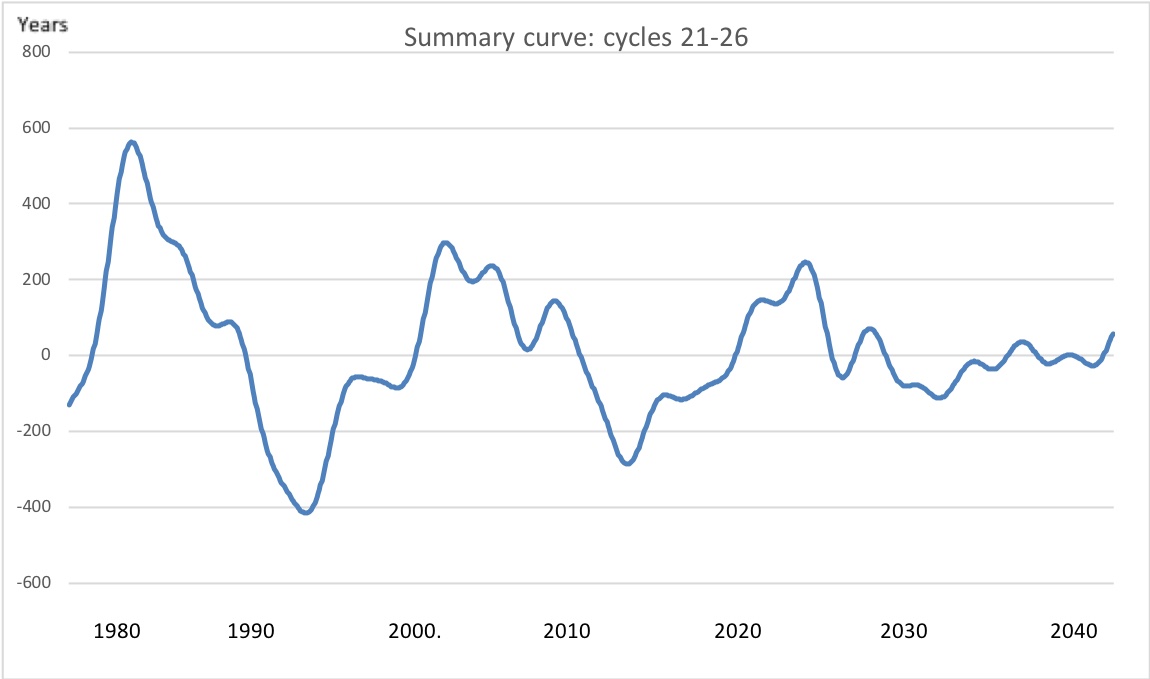} 
\includegraphics[width=35pc]{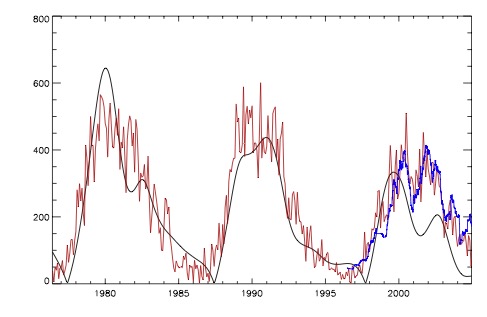}  
\caption{Top plot: The summary curve of EVs, or PCs, derived from the data for cycles 21-23 and extrapolated to cycles 24, 25 and 26 (taken from \citep{Zharkova2015, Zharkova2020a}). Bottom plot:  Modulus summary curve, derived from the summary curve above for cycles 21-23 overplotted on the averaged sunspot numbers used as the current solar activity index  (taken from \citep{Zharkova2015})  confirming the summary curve of EVs as an additional proxy of solar activity \citep[see also the recent verification for cycles 21-24 in ][]{Zharkova2022}. } 
\label{sa_sbmf} 
\end{figure}  

 This summary curve of these two PCs derived for cycles 21-23 and predicted for cycle 24-26 is plotted in  Fig. \ref{sa_sbmf} (top plot) also showing the variations of the dominant solar background magnetic field: northern for positive and southern for negative amplitudes.  The prediction of the summary curve to cycles 25 and 26 presented in Fig. \ref{sa_sbmf}  (top plot) taken from Fig. 2, bottom plot in \citep{Zharkova2015}) shows a noticeable decrease of the predicted average sunspot numbers in cycle 25 to $\approx 80\%$ of that in cycle 24 and in cycle 26 to $\approx 40\%$ that is linked to a reduction of the amplitudes and an increase of phases of the principal components of SBMF.  The prediction of the summary curve by thousand years backward and forward  presented in Fig.3 \citep{Zharkova2015} revealed the occurrence of grand solar cycle of 350-400 years, reproducing the well-known Maunder, Wolf, Oort, Homeric and many other Grand Solar Minima.

This summary curve was proposed  \citep{Zharkova2015} as a new solar activity proxy since  the module of the summary curve fits rather closely the averaged sunspot numbers currently used as a solar activity index (Fig. \ref{sa_sbmf}, bottom plot). 
A remarkable resemblance between the modulus summary curve and the curves describing  the averaged smoothed sunspot numbers or the averaged sunspot magnetic flux in cycles 21-23 with some small exception for the descending phase of cycle 23, which was later explained by the strongly inflated sunspot numbers used at Locarno observatory \citep{Clette2014}.  After their  correction, the averaged sunspot numbers in cycle 23 fit rather closely the modulus summary curve presented in Fig. \ref{sa_sbmf}. 

 Hence, from the one hand, this modulus summary curve is found to be a good proxy of the traditional solar activity contained in the averaged sunspot numbers. On the other hand, this summary curve is a derivative from the principal components of SBMF with clear mathematical functionalities representing at the same time the real physical processes - poloidal field dynamo waves - generated by the solar dynamo \citep{Zharkova2015}. Recently, the other PCs, or eigen vectors of SBMF derived from the SBMF data for four cycle 21-24 supposedly produced by quadruple, sextuple and octuple magnetic sources are shown linked to various  flare activity indices in  soft X-ray and radio emission \citep{Zharkova2022}. 
 
The suggestion of usage of the summary curve of two eigen vectors, or principle components of the solar background magnetic field as a new proxy of solar activity has been recently supported by other predictions of the solar activity in cycle 25  \citep{Kitiashvili2020, Obridko2021} obtained from the same solar magnetic field components measured from the same WSO magnetic synoptic maps as those reported earlier\citep{shepherd14, Zharkova2015}. 
 
 \subsubsection{Links of volcanic eruptions with the SBMF variations}\label {sbmf_res}
\begin{figure}[h]
\includegraphics[width=40pc]{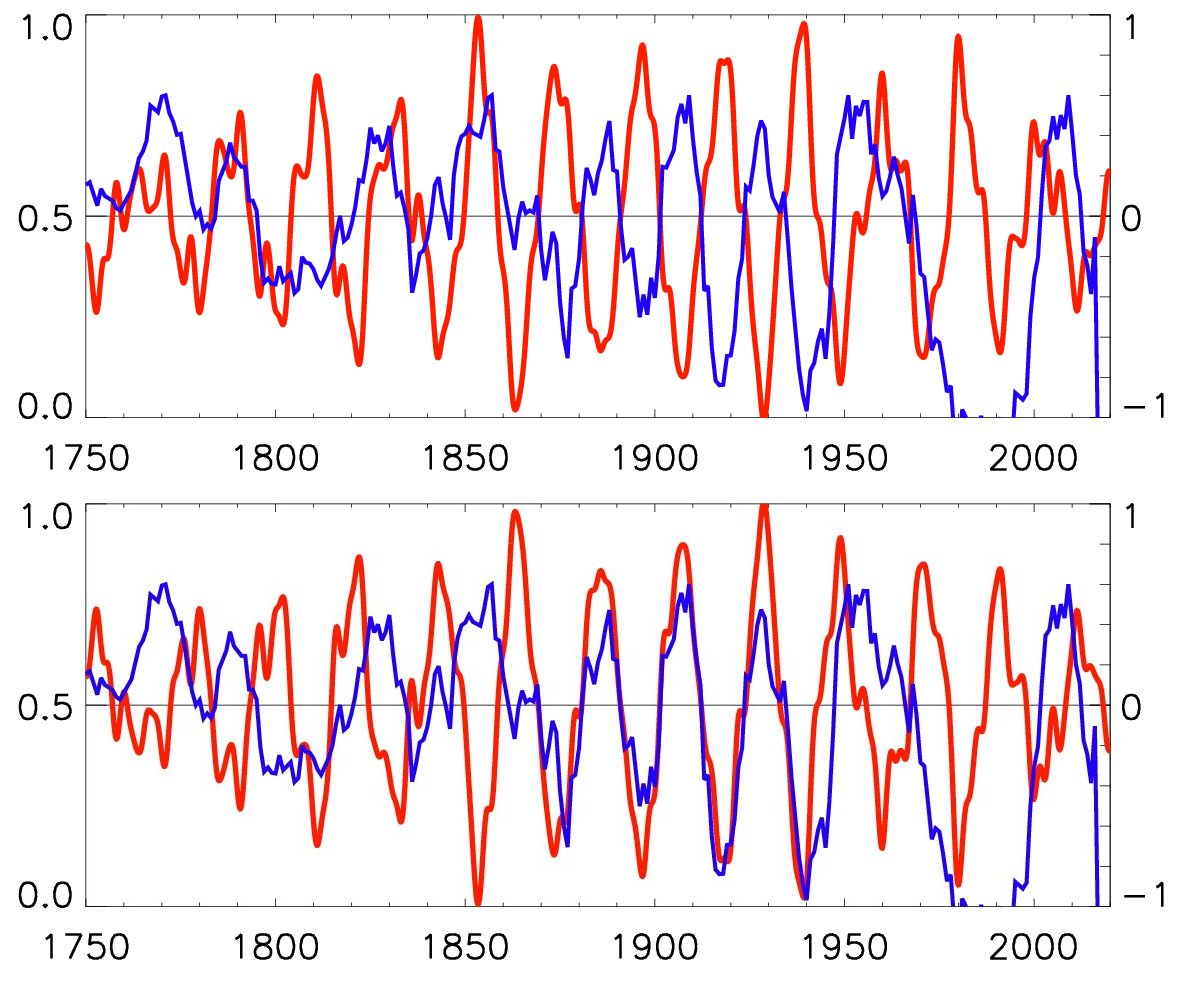} 
\caption{Top plot:  The summary curve of eigen vectors (EV) of the solar background magnetic field  \citep{Zharkova2015}  (red curve) normalised  by its maximum (the right Y-axis)  versus the averaged normalised number of volcanic eruptions (VEs) (blue curve) (the left Y-axis). Positive magnitudes of the summary  curve correspond to the northern polarity and negative ones to the southern polarity of SBMF.  Bottom plot: the volcanic eruption (VE) numbers (left Y-axis, blue line)  versus the inverted summary  curve of eigen vectors (EV1) (the right Y-axis, red line)  with positive magnitudes corresponding to southern polarity and negative to the northern one. For the correlation coefficients between these two curves.see the text and  Fig.\ref{scat1_ve_ev}. } 
\label{volc_sbmf} 
\end{figure}

Let us compare the temporal variations of the frequency of volcanic eruptions (VE) and variations of the summary curve of the PCs, or the eigen vectors, of the SBMF \citep{Zharkova2015} defining the solar activity during the solar cycles  from 1750 to 2020.  In order to compare the frequencies of volcanic eruptions with variations of the summary curve of the PCs of SBMF, the arbitrary amplitudes of the summary curve variations were normalised  by its maximal magnitude. The comparison of these two series with the frequencies of volcanic eruptions inverted from minima on the top to maxima at the bottom is shown in Fig. \ref{volc_sbmf}  for a symbolic 11 year cycle defined by averaged sunspot numbers (left plot) and for symbolic  double solar cycle defined by two 11 cycles of the summary curve of eigen vectors (EVs) of SBMF, one cycle with the northern and another with southern magnetic polarity (right plot).

  Volcanic eruption (VE) frequencies  versus a symbolic 22 year solar cycle (two cosine functions (yellow line) with with the opposite polarities as shown in Fig. \ref{sa_sbmf}, bottom plot), defined by a summary curve of eigen vectors (EVs) of SBMF (Fig. \ref{sa_sbmf}, top plot). The summary VE frequencies were calculated for the intervals marked by rectangular, for the EVs  cycles showing leading magnetic polarities (northern  by a positive sign and southern by a negative one) for the two periods: 1750-1868 (black line) and 1868-1950 \&1990-2020 (indigo line).
 
It the period 1868 - 1950 and 1990-2020 the frequency of volcanic eruptions reveals a clear pattern of its maxima following the maxima or the descending phase of the magnetic field curve with the southern polarity while minima fit rather close the maxima or ascending phase of the magnetic field maxima with the northern polarity.  While  in the earlier years (1762-1868) and in the years of 1950-1980 this link disappeared.  Hence, the volcanic eruption frequencies are found following the solar magnetic field variations with a smaller degree of correlation for 8 cycles before 1868 and with much higher correlations for 11 cycles in 1868- 1950 and 1990-2020.

\begin{figure}[h]
\includegraphics[width=40pc]{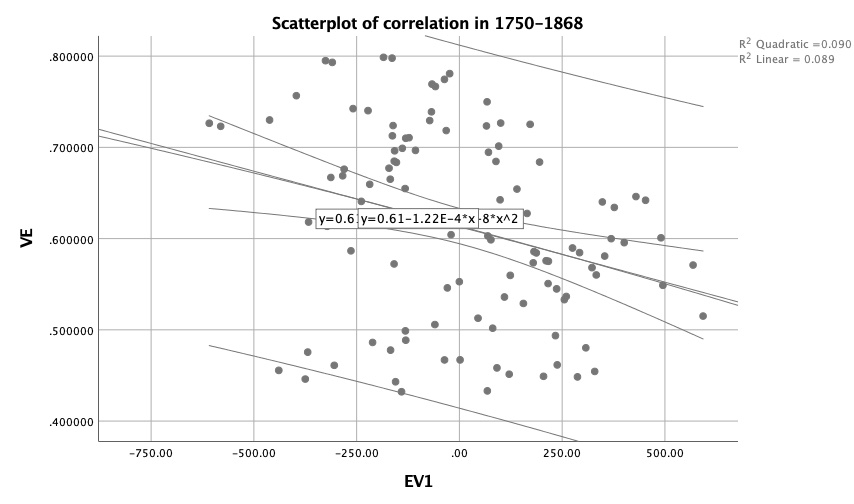} 
\includegraphics[width=40pc]{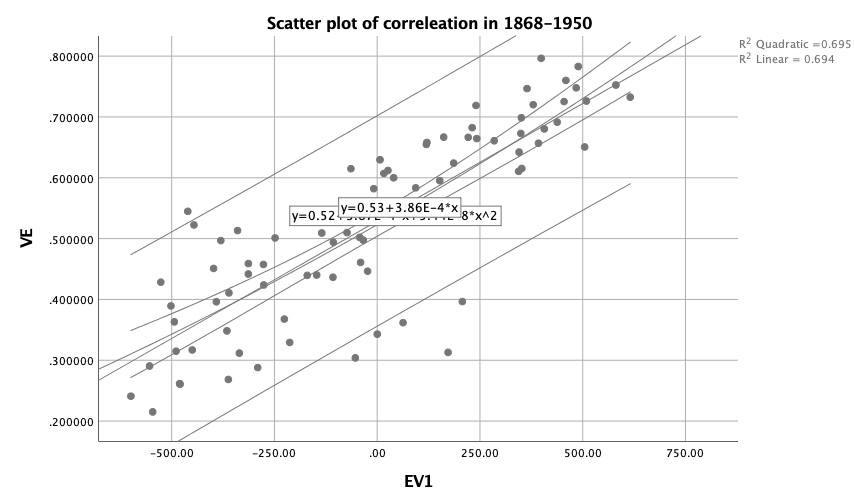} 
\caption{ Top plot: The scatter plot of the correlation of  the inverted volcanic frequency (VE) with the inverted eigen vectors (EV1) of SBMF for the data from 1750-1868. Nonparametric (Spearman) correlation coefficient is -0.325.  Bottom plot: The scatter plot of the correlation of  volcanic frequency (VE) with the inverted eigen vectors (EV1) of SBMF for the data from 1868-1950. The Spearman correlation coefficient is 0.840, which defines a strong correlation of VE with the EV of the southern polarity.  The central lines provide the best linear and quadratic fits of correlation coefficients, the near lines define standard deviations of the fits and the outer lines  show the 95$\%$ confidence intervals for the derived correlation coefficients.} 
\label{scat1_ve_ev} 
\end{figure}

 The comparison of this normalised summary curve of eigen vectors (EV) with the whole curve of volcanic eruption (VE) frequencies  is plotted in Fig. \ref{volc_sbmf}  for real EV of SBMF and VE (top plot) and for the inverted EV, curve EV1, (bottom plot). In EV1 the maxima correspond to the EV cycles with southern polarity and coincide with maxima in VEs and minima correspond to  the EV cycles with northern polarity that happen to coincide with the minima in VEs  as shown in Fig. \ref{volc_sbmf}, top plot. The correlation of the overall VE and EV1, or the inverted EV data was calculated using the IBM SPSS Statistics package (v25) and produced  the positive correlation coefficient of 0.21.  Since there were two distinct temporal intervals where the curves in Fig. \ref{volc_sbmf} show different properties: 1762-1868 and 1868-1950,  we also looked at the correlation coefficients of these series "by parts" for these two periods and  the correlation results are presented in the scatter plots in Fig. \ref{scat1_ve_ev}  for 1750-1868  (top plot) and for 1868-1950 (bottom plot). The disappearance of a link of VEs with the solar or terrestrial magnetic fields in 1950s to 1980s can be linked to the open nuclear bomb testing as the link is restored  a few decades after this testing was banned, so this period was omitted from consideration.

 The correlation coefficient of VEs with EV1, or the inverted EVs, was -0.33  for the period of volcanic eruptions in 1750-1868, while reaching the magnitude of 0.84  for the volcanic eruptions in the period of 1868-1950. In summary, there is  a strong correlation of volcanic eruptions  VEs with EVs of the southern polarity seen in  Fig. \ref{scat1_ve_ev} following the links between VE and EV demonstrated in Fig. \ref{volc_sbmf} (top plot).  The linear and quadratic fits of correlation coefficients in the scatter plots are shown in Fig. \ref{scat1_ve_ev} revealing no essential differences between these fits. 

The standard deviations (STDs) from the fits are shown by the inner set of lines about the fits, while the confidence intervals for a 95$\%$ confidence level are shown by the outer lines for the both data sets revealing very close fit of the correlation coefficient in the second period (1868-1950) and more scattered one in the first one (1750-1868). 
Hence, the volcanic eruption  (VE) frequencies have maxima every 22 years during the periods when the summary curves  of eigen vectors (EVs) have the southern polarity. This coincides with the period of 22 years found  from the volcanic  eruption frequencies  in the wavelet section using the Morlet wavelet shown in Fig. \ref{wavelet}.  

 Also the total numbers of volcanic eruptions  shown in Fig. \ref{wolf_num} (bottom right plot) during a 22 year cycle (red line) for the period of 1750-1868 (black curve) and 1868-1950 (indigo curve) confirm that the maximum volcanic eruptions occur during the maxima of solar activity cycles of EVs of SBMF with southern magnetic polarity. This finding is in line  with the other studies showing  the high speed solar wind streams or energetic proton fluxes  to be the drivers of the most powerful earthquakes \citep{Odintsov2006, Marchitelli2020a, Marchitelli2020b}.

 Furthermore,  a strong correlation of the frequencies of volcanic eruptions in 1868-1950  and solar magnetic cycles with the southern magnetic polarity can be understood in the terms of accepted views that the increase of geomagnetic disturbances often correspond to an increase in the interplanetary magnetic field of the southern polarity (see, for example, \citep{Maezawa1974, Perreault1978, Stauning1994, Stauning1995, Gonzales1999, Prosovetsky2011}).   A  possible reason for the lower correlation in the early years of 1750-1868 is suggested in section below. 
 
 \section{Discussion and conclusions} \label{discussion}

 \subsection{Volcanic eruptions and motion of the North pole} \label{n_pole}

The Carrington solar event in 1859, the largest recorded solar magnetic event, has been associated with the external field changes with the minimum -1760 nT at the Colaba magnetic Observatory in Bombay \citep{Cliver2013,  Mandea2020}.  Also, there is the evidence that a geomagnetic jerk has occurred around 1860 \citep{Newitt2002, Newitt2007}. The geomagnetic jerk is a relatively abrupt change in the rate of secular variations in one or more parameters of the Earth's magnetic field. One of the most powerful geomagnetic jerks was observed during of 1969-1970. Until about 1971, the northern magnetic pole moved more or less uniformly at a speed of about 10 km/year, then suddenly began to accelerate. This acceleration of the pole motion is associated with the so-called geomagnetic jerk that occurred in 1969-1970 \citep{Newitt2002, Newitt2007}. 

\begin{figure}[ht]
\center{\includegraphics[width=30pc]{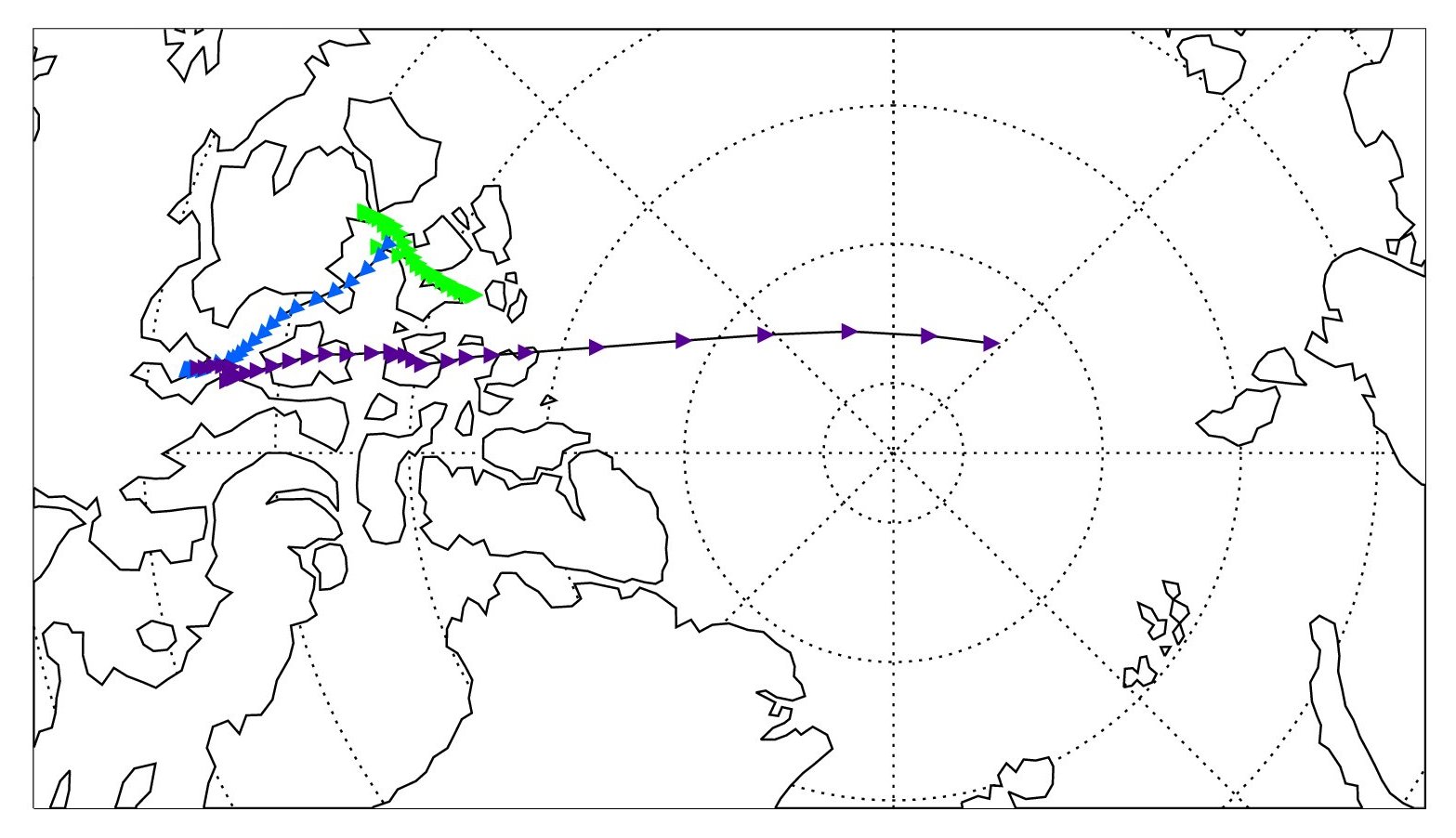}}
\caption{Reconstructed locations of the north magnetic pole from 1500 to 2020 \citep{NGDC2021}.}
\label{pole} 
\end{figure}

The geomagnetic jerk, which took place around 1860, may help to explain the change in direction of the northern magnetic pole, which was suggested when considering earthquakes \citep{USGS2019}. 
Another possible confirmation of the internal restructuring of the Earth in the 60s of the 19th century may be a sharp change in the direction of the motion of the Earth's magnetic pole (Fig. \ref{pole}) according to model calculations \citep{NGDC2021}. 

In 1760-1860, the magnetic pole was moving away (see Fig. \ref{pole}), and after 1861 it began returning back by rapidly approaching the geographic pole of the Earth. This position was approached as close as possible in 2018, and the magnetic pole began to move away again. These dates coincide very well with the periods identified by us from a comparison of the number of volcanic eruptions and the SBMF variations.  

\subsection{Volcanic eruptions follow a 22 year magnetic field cycle}
In this paper we explored the frequencies of volcanic eruptions  in the past two centuries and their possible links to the variations of solar activity and solar background magnetic field.  Contrary to \citep{Stothers1989},  we obtained a dominant ~ 22-year period of volcanic activity (VEI $\ge 2$, 3829 eruptions) and a much weaker peak for a period of 10.7 years. More volcanic eruptions  are found to occur during maximum phases of the doubled solar cycles of 22 years of SBMF when the solar background magnetic field  has a southern polarity.

We established a rather high positive correlation of the number of volcanic eruptions with the solar background magnetic field of southern polarity during the period of 1868-1950  and  much smaller negative correlation with the southern polarity magnetic fields during the period 1750-1868. The strong correlation recorded in 8 cycles of SBMF after 1868 suggests that possible physical mechanisms of volcanic eruptions are linked to the increased solar activity features, e.g. the solar background (poloidal) magnetic field of southern polarity, energetic particles of solar wind and solar flares and large-scale shocks  produced by interacting solar magnetic loops of the toroidal field. 

This link of volcanic eruptions with the solar background  magnetic field of southern polarity is likely to include  the increased electro-magnetic interaction of the SBMF with the terrestrial magnetic field causing geomagnetic disturbances \citep{Maezawa1974, Perreault1978, Stauning1994, Stauning1995, Gonzales1999} that can lead to shifts of the crusts, most powerful earthquakes \citep{Odintsov2006} and volcanic eruptions.  This can be  also influenced by the perturbations of energetic particles and waves on the Earth's air masses \citep{Stothers1989} or an increase in precipitation \citep{Stothers1989} that can lead to an increase in the tectonic /volcanic instability.  

The possible reasons of a reduced correlation between volcanic eruptions and solar activity index recorded in the period of 1750-1868 can be associated with  the geomagnetic jerk and related migration of the North pole towards lower latitudes.  Although, some of the differences between the distributions of eruption frequencies over the phases of solar cycles of 11 years can be related to the differences in solar activity indices defined by sunspots and eigen vectors of magnetic fields related to different types of solar magnetic fields: toroidal and poloidal, discussed in the forthcoming paper.  While  the lack of correlation in 1950-1980 can be linked to the open air nuclear bomb testings that distorted the effects of the solar magnetic field.  

In summary, the increase of volcanic eruptions established during the solar cycles with the southern polarity of SBMF emphasises the importance of solar-terrestrial interaction in volcanic eruptions. This link can also play an important role in the next few decades affected by the modern Grand Solar Minimum (2020-2053) \citep{Zharkova2015, Zharkova2020a} because in cycle 26 the summary EV of the solar background magnetic field will have the southern polarity. Hence, despite links between volcano occurrences and solar activity not being clear yet, the consequences of volcanic eruptions for the terrestrial atmosphere during cycle 26 with the southern polarity of solar magnetic field can be expected to be noticeable during the modern GSM (2020-2053).

\section{Acknowledgments}
 The authors express their deepest gratitude to the anonymous referee for useful and constructive comments from which the paper strongly benefited. The authors wish to express their many thanks to the Smithsonian Institution staff for providing the access to the GVP database containing the volcanic eruption data  and the Solar Influences Data Analysis Center (SIDC) at the Royal Observatory of Belgium for providing the averaged sunspot numbers. The authors also express their deepest gratitude to the staff and directorate of Wilcox Solar Observatory, Stanford, US, for providing the coherent long-term observations of full disk synoptic maps of the solar background magnetic field. 

\section*{Author contributions statement}

I.V. gathered and processed the volcanic frequency data while V.Z. provided and analysed the solar background magnetic field data.  V.Z. and I.V. compared and analysed the results, wrote and reviewed the manuscript. 

\section*{Additional information}

The authors do not have any competing financial interests. 

\bibliographystyle{aa}
\bibliography{volcanos1}

\begin{thebibliography}{64}
\expandafter\ifx\csname natexlab\endcsname\relax\def\natexlab#1{#1}\fi

\bibitem[{USG(2021)}]{USGS2019}
 2021, {USGS}, {U}.{S}. {G}eological {S}urvey. {E}arthquake {H}azards
  {P}rogram, accessed June 7, 2019

\bibitem[{NGD(2021)}]{NGDC2021}
 2021, {NGDC}, The {N}ational {G}eophysical {D}ata {C}enter. {W}andering of the
  {G}eomagnetic poles, accessed July 23, 2021

\bibitem[{{Akasofu}(2010)}]{Akasofu2010}
{Akasofu}, S.-I. 2010, Natural Science, 2, 1211

\bibitem[{{Anderson}(1974)}]{Anderson1974}
{Anderson}, D.~L. 1974, Science, 186, 49

\bibitem[{{Bumba}(1986)}]{Bumba1986}
{Bumba}, V. 1986, Contributions of the Astronomical Observatory Skalnate Pleso,
  15, 495

\bibitem[{{Casati}(2014)}]{Casati2014}
{Casati}, M. 2014, in EGU General Assembly Conference Abstracts, EGU General
  Assembly Conference Abstracts, 1385

\bibitem[{{Clette} {et~al.}(2014){Clette}, {Svalgaard}, {Vaquero}, \&
  {Cliver}}]{Clette2014}
{Clette}, F., {Svalgaard}, L., {Vaquero}, J.~M., \& {Cliver}, E.~W. 2014, Space
  Science Reviews, 186, 35

\bibitem[{{Cliver} \& {Dietrich}(2013)}]{Cliver2013}
{Cliver}, E.~W. \& {Dietrich}, W.~F. 2013, Journal of Space Weather and Space
  Climate, 3, A31

\bibitem[{{De Marchi}(1911)}]{DeMarchi1895}
{De Marchi}, L. 1911, Scientia, 5, 310

\bibitem[{{Easterbrook}(2016)}]{Easterbrook2016}
{Easterbrook}, D.~J. 2016, {Evidence-based Climate Science} (Elsevier)

\bibitem[{{Eddy}(1976)}]{Eddy1976}
{Eddy}, J.~A. 1976, Science, 192, 1189

\bibitem[{Global Volcanism~Program(2013)}]{Global2013}
Global Volcanism~Program, G. 2013, {Volcanoes of the World, v. 4.10.0 (14 May
  2021)}, downloaded June 9, 2021

\bibitem[{{Gonzalez} {et~al.}(1999){Gonzalez}, {Tsurutani}, \& {Cl{\'u}a de
  Gonzalez}}]{Gonzales1999}
{Gonzalez}, W.~D., {Tsurutani}, B.~T., \& {Cl{\'u}a de Gonzalez}, A.~L. 1999,
  Space Sci. Reviews, 88, 529

\bibitem[{Gray {et~al.}(2010)Gray, Beer, Geller, Haigh, Lockwood, Matthes,
  Cubasch, Fleitmann, Harrison, Hood, Luterbacher, Meehl, Shindell, van Geel,
  \& White}]{Gray2010}
Gray, L.~J., Beer, J., Geller, M., {et~al.} 2010, Reviews of Geophysics, 48

\bibitem[{{Han} {et~al.}(2004){Han}, {Guo}, \& {Ma}}]{Han2004}
{Han}, Y., {Guo}, J., \& {Ma}, C. 2004, Sci. China Ser. B: Phys. Mech. Astron,
  47, 173

\bibitem[{{Herdiwijaya} {et~al.}(2014/10){Herdiwijaya}, Johan, \&
  {Nurzaman}}]{Herdiwijaya2014}
{Herdiwijaya}, D., Johan, A., \& {Nurzaman}, M.~Z. 2014/10, in Proceedings of
  the 2014 International Conference on Physics (Atlantis Press), 105--108

\bibitem[{{Jensen}(1902)}]{Jensen1902}
{Jensen}, H.~I. 1902, Journal and Proceedings of the Royal Society of New South
  Wales, 36, 42

\bibitem[{{Kelly}(1996)}]{Kelly1996}
{Kelly}, A.~C. 1996, in ESA Special Publication, Vol.~1, ESA Special
  Publication, 106--113

\bibitem[{{Kitiashvili}(2020)}]{Kitiashvili2020}
{Kitiashvili}, I.~N. 2020, Astrophysical J., 890, 36

\bibitem[{Kluge(1863)}]{Kluge1863}
Kluge, E. 1863, {Ueber Synchronismus und Antagonismus yon vulkanischen
  Eruptionen und die Beziehungen derselben zu den Sonnenfiecken und
  erdmagnetischen Variationen} (Leipzig: Engelmann)

\bibitem[{{K\"{o}ppen}(1896)}]{Koppen1896}
{K\"{o}ppen}, W. 1896, Himmel und Erde, 8, 529

\bibitem[{{Lean} {et~al.}(1995){Lean}, {Beer}, \& {Bradley}}]{Lean1995}
{Lean}, J., {Beer}, J., \& {Bradley}, R. 1995, Geophys. Res. Letters, 22, 3195

\bibitem[{{Lockwood} \& {Owens}(2014)}]{Lockwood2014}
{Lockwood}, M. \& {Owens}, M.~J. 2014, Astrophys. J. Letters, 781, L7

\bibitem[{{Lockwood} {et~al.}(1999){Lockwood}, {Stamper}, \&
  {Wild}}]{Lockwood1999}
{Lockwood}, M., {Stamper}, R., \& {Wild}, M.~N. 1999, Nature, 399, 437

\bibitem[{{Love} \& {Thomas}(2013)}]{Love2013}
{Love}, J.~J. \& {Thomas}, J.~N. 2013, Geophys. Res. Letters, 40, 1165

\bibitem[{{Lyons}(1899)}]{Lyons1899}
{Lyons}, C.~J. 1899, Monthly Weather Review, 27, 144

\bibitem[{{Ma} {et~al.}(2018){Ma}, {Yin}, \& {Han}}]{Ma2018}
{Ma}, L., {Yin}, Z., \& {Han}, Y. 2018, Earth Science Research, 7, 110

\bibitem[{{Maezawa}(1974)}]{Maezawa1974}
{Maezawa}, K. 1974, Planetary and Space Science J., 22, 1443

\bibitem[{Mandea \& Chambodut(2020)}]{Mandea2020}
Mandea, M. \& Chambodut, A. 2020, Surveys in Geophysics, 41

\bibitem[{{Martichelli} {et~al.}(2020{\natexlab{a}}){Martichelli},
  {Harabaglia}, {Troise}, \& {De Natale}}]{Marchitelli2020a}
{Martichelli}, V., {Harabaglia}, P., {Troise}, C., \& {De Natale}, G.
  2020{\natexlab{a}}, Scientific Reports, 10, 11495

\bibitem[{{Martichelli} {et~al.}(2020{\natexlab{b}}){Martichelli},
  {Harabaglia}, {Troise}, \& {De Natale}}]{Marchitelli2020b}
{Martichelli}, V., {Harabaglia}, P., {Troise}, C., \& {De Natale}, G.
  2020{\natexlab{b}}, Front.Earth Sci, 10, , doi =
  {doi:10.3389/feart.2020.595209}

\bibitem[{{Mazzarella} \& {Palumbo}(1989)}]{Mazzarella_Palumbo1989}
{Mazzarella}, A. \& {Palumbo}, A. 1989, J. Volcanology and Geothermal Research,
  39, 89

\bibitem[{Newhall \& Self(1982)}]{Newhall_Self1982}
Newhall, C. \& Self, S. 1982, J. Geophys. Research, 87, 1231

\bibitem[{Newitt \& Dawson(2007)}]{Newitt2007}
Newitt, L. \& Dawson, E. 2007, Geophys. J. of the Royal Astronomical Society,
  78, 277

\bibitem[{Newitt {et~al.}(2002)Newitt, Mandea, McKee, \& Orgeval}]{Newitt2002}
Newitt, L., Mandea, M., McKee, L., \& Orgeval, J.-J. 2002, EOS Transactions,
  83, 381

\bibitem[{{Obridko} {et~al.}(2021){Obridko}, {Sokoloff}, {Pipin}, {Shibalvaa},
  \& {Livshits}}]{Obridko2021}
{Obridko}, V.~N., {Sokoloff}, D.~D., {Pipin}, V.~V., {Shibalvaa}, A.~S., \&
  {Livshits}, I.~M. 2021, Monthly Notices of Royal Astr. Soc, 4990

\bibitem[{{Odintsov} {et~al.}(2006){Odintsov}, {Boyarchuk}, {Georgieva}, \&
  {Atanasov}}]{Odintsov2006}
{Odintsov}, S., {Boyarchuk}, K., {Georgieva}, K., \& {Atanasov}, D. 2006, J.
  Physics and Chemistry of the Earth, 31, 88

\bibitem[{{O'Reilly}(1898)}]{O'Reilly1898}
{O'Reilly}, J.~P. 1898, Proceedings of the Royal Irish Academy (1889-1901), 5,
  392

\bibitem[{Parker {et~al.}(1994)Parker, Jones, Folland, \& Bevan}]{Parker1994}
Parker, D.~E., Jones, P.~D., Folland, C.~K., \& Bevan, A. 1994, J. Geophys.
  Research: Atmospheres, 99, 14373

\bibitem[{{Perreault} \& {Akasofu}(1978)}]{Perreault1978}
{Perreault}, P. \& {Akasofu}, S.~I. 1978, Geophys. Journal, 54, 547

\bibitem[{Prosovetsky \& Myagkova(2011)}]{Prosovetsky2011}
Prosovetsky, D. \& Myagkova, I. 2011, Geomagnetism and Aeronomy, 51

\bibitem[{{Sapper}(1930)}]{Sapper1930}
{Sapper}, K. 1930, Volcano Letters, 302, 2

\bibitem[{{Schmidt} \& {Lipson}(2009)}]{Schmidt09}
{Schmidt}, M. \& {Lipson}, H. 2009, Science, 324, 81

\bibitem[{{Schneider} \& {Mass}(1975)}]{Schneider1975}
{Schneider}, S.~H. \& {Mass}, C. 1975, Science, 190, 741

\bibitem[{{Shepherd} {et~al.}(2014){Shepherd}, {Zharkov}, \&
  {Zharkova}}]{shepherd14}
{Shepherd}, S.~J., {Zharkov}, S.~I., \& {Zharkova}, V.~V. 2014, Astrophysical
  J., 795, 46

\bibitem[{{{SILSO} World Data Center}(2021)}]{SILSO}
{{SILSO} World Data Center}. 2021, International Sunspot Number Monthly
  Bulletin and online catalogue

\bibitem[{{Sobolev} \& {Demin}(1980)}]{Sobolev1980}
{Sobolev}, N.~V. \& {Demin}, V.~M. 1980, {Mechano-electrical Phenomena in the
  Earth} ({Nauka}), 215pp

\bibitem[{{Stamper} {et~al.}(1999){Stamper}, {Lockwood}, {Wild}, \&
  {Clark}}]{Stamper1999}
{Stamper}, R., {Lockwood}, M., {Wild}, M.~N., \& {Clark}, T.~D.~G. 1999, J.
  Geophys. Research, 104, 28325

\bibitem[{{Stauning}(1994)}]{Stauning1994}
{Stauning}, P. 1994, J. Geophys. Research, 99, 17309

\bibitem[{{Stauning} {et~al.}(1995){Stauning}, {Clauer}, {Rosenberg},
  {Friis-Christensen}, \& {Sitar}}]{Stauning1995}
{Stauning}, P., {Clauer}, C.~R., {Rosenberg}, T.~J., {Friis-Christensen}, E.,
  \& {Sitar}, R. 1995, J. Geophys. Research, 100, 7567

\bibitem[{{Steinhilber} {et~al.}(2012){Steinhilber}, {Abreu}, {Beer},
  {Brunner}, {Christl}, {Fischer}, {Heikkila}, {Kubik}, {Mann}, {McCracken},
  {Miller}, {Miyahara}, {Oerter}, \& {Wilhelms}}]{Steinhilber12}
{Steinhilber}, F., {Abreu}, J.~A., {Beer}, J., {et~al.} 2012, Proceedings of
  the National Academy of Science, 109, 5967

\bibitem[{{Steinhilber} {et~al.}(2009){Steinhilber}, {Beer}, \&
  {Fr{\"o}hlich}}]{Steinhilber09}
{Steinhilber}, F., {Beer}, J., \& {Fr{\"o}hlich}, C. 2009, Geophys. Res.
  Letters, 36, L19704

\bibitem[{Stothers(1989)}]{Stothers1989}
Stothers, R.~B. 1989, J. Geophys. Research, 94, 17371

\bibitem[{{St{\v{r}}e{\v{s}}tik}(2003)}]{Strestik2003}
{St{\v{r}}e{\v{s}}tik}, J. 2003, in ESA Special Publication, Vol. 535, Solar
  Variability as an Input to the Earth's Environment, ed. A.~{Wilson}, 393--396

\bibitem[{Vasilyeva(2020)}]{Vasilyeva2020}
Vasilyeva, I. 2020, Space Science and Technology (Ukraine), 26, 90

\bibitem[{{Velasco Herrera} {et~al.}(2021){Velasco Herrera}, {Soon}, \&
  {Legates}}]{velasco2021}
{Velasco Herrera}, V.~M., {Soon}, W., \& {Legates}, D.~R. 2021, Advances in
  Space Research, 68, 1485

\bibitem[{{Wolf}(1870)}]{Wolf1870}
{Wolf}, R. 1870, Monthly Notices of Royal Astron.Soc., 30, 157

\bibitem[{Zharkova(2020)}]{Zharkova2020a}
Zharkova, V. 2020, Editorial paper, Temperature, 7, 217

\bibitem[{{Zharkova}(2021)}]{Zharkova2021}
{Zharkova}, V. 2021, chapter in a book "Solar system planets and exoplanets",
  30 pp.

\bibitem[{{Zharkova} \& {Shepherd}(2022)}]{Zharkova2022}
{Zharkova}, V.~V. \& {Shepherd}, S.~J. 2022, Monthly Notices of Royal
  Astron.Soc., 512, 5085

\bibitem[{{Zharkova} {et~al.}(2015){Zharkova}, {Shepherd}, {Popova}, \&
  {Zharkov}}]{Zharkova2015}
{Zharkova}, V.~V., {Shepherd}, S.~J., {Popova}, E., \& {Zharkov}, S.~I. 2015,
  Scientific Reports, 5, 15689

\bibitem[{{Zharkova} {et~al.}(2012){Zharkova}, {Shepherd}, \&
  {Zharkov}}]{Zharkova12}
{Zharkova}, V.~V., {Shepherd}, S.~J., \& {Zharkov}, S.~I. 2012, Monthly Notices
  of Royal Astron.Soc., 424, 2943

\bibitem[{{Zharkova} {et~al.}(2019){Zharkova}, {Shepherd}, {Zharkov}, \&
  {Popova}}]{Zharkova2019}
{Zharkova}, V.~V., {Shepherd}, S.~J., {Zharkov}, S.~I., \& {Popova}, E. 2019,
  Scientific Reports, 9, 9197

\bibitem[{{Zharkova} {et~al.}(2022){Zharkova}, {Vasilieva}, {Shepherd}, \&
  {Popova}}]{Zharkova2022b}
{Zharkova}, V.~V., {Vasilieva}, I., {Shepherd}, S.~J., \& {Popova}, E. 2022,
  Solar Physics., subm, 24

\end{thebibliography}

\end{document}